\documentclass[%
reprint,
superscriptaddress,
 amsmath,amssymb,
 aps,
]{revtex4-1}

\usepackage{graphicx}
\usepackage{dcolumn}
\usepackage{bm}
\usepackage[mathlines]{lineno}
\usepackage[colorlinks=true]{hyperref}
\usepackage{mathrsfs}
\usepackage{xcolor}
\usepackage{float}
\usepackage{ulem}
\usepackage[utf8]{inputenc}

\newcommand{\be}{\begin{equation}}
\newcommand{\ee}{\end{equation}}
\newcommand{\bea}{\setlength\arraycolsep{2pt} \begin{eqnarray}}
\newcommand{\eea}{\end{eqnarray}}

\begin{document}


\title{Information paradox and corrected thermodynamics for black holes}
\author{Ming Zhang}
\email{mingzhang@jxnu.edu.cn}
\affiliation{College of Physics and Communication Electronics, Jiangxi Normal University, Nanchang 330022, China}
\affiliation{Department of Physics, Beijing Normal University, Beijing 100875, China}

\date{\today}

\begin{abstract}
We generally consider that entropy and temperature of a spherically symmetric black hole are corrected by quantum effect. We calculate interior entropy variation of massless scalar field and compare it with corrected Bekenstein-Hawking entropy variation for the evaporating black hole. We find that the corrected ratio between them is greater than the uncorrected one. The information paradox is then discussed.
\end{abstract}


\maketitle


\section{Introduction}
Naively speaking, it seems nonsense to talk about the volume of a black hole, as the time and space are interchanged inside it. However, the possibility of defining the volume was shown in case of a somewhere-timelike Killing vector admitting by the spacetime \cite{Parikh:2005qs}. The volume defined in way of choosing stationary time slices is independent of time. The definitions of Kodama volume \cite{Hayward:1997jp,Hayward:1999ek}, vector volume \cite{Ballik:2013uia} for black holes are also intriguing and heuristic.  In spacetime charaterized with a varying cosmological constant, a geometric volume \cite{Kastor:2009wy}, which is formally equivalent to Parikh volume \cite{Parikh:2005qs}, can be defined. This kind of volume plays as a role conjugated to the pressure defined by the varying cosmological constant. This not only makes the thermodynamic first law be consistent with the Smarr formula \cite{Chamblin:1999hg}, but also provides new insights into the study of the black hole phase transition \cite{Kubiznak:2012wp}.

In the flat Minkoswki spacetime, the volume inside a $n$-dimensional sphere $\mathcal{S}$ can be viewed as a $(n+1)$-dimensional spacelike surface $\Sigma$ bounded by $\mathcal{S}$. The surface $\Sigma$ lies on the same simultaneity surface $\mathcal{S}$ \cite{Christodoulou:2014yia}. It is not difficult to prove that the  simultaneity surface is also the largest spherically symmetric surface bounded by $\mathcal{S}$ in the flat case. However, in the curved spacetime, as it is difficult to define the simultaneity surface,  the volume of a spherically symmetric black hole is suggested to be the largest spherically symmetric surface bounded by $\mathcal{S}$ \cite{Christodoulou:2014yia}. 

Along this way, the Christodoulou-Rovelli (CR) volume of a spherically symmetric black hole was defined \cite{Christodoulou:2014yia,Ong:2015tua,Ong:2015dja,Bhaumik:2016sav,Wang:2017zfn,Zhang:2019pzd}. The volume of the spinning Kerr black hole was also raised \cite{Bengtsson:2015zda}. The volume of the collapsed object is proved to be increasing with time, which to some extent can be used to explain the paradox of the information conservation \cite{Nikolic:2018ith}. Based on this concept of black volume, statistical Boltzmann entropy of massless scalar field inside a black hole was firstly raised in \cite{Zhang:2015gda} and then elaborated in \cite{Majhi:2017tab,Zhang:2017aqf,Wang:2018dvo,Yang:2018arj,Ali:2018sqk,Han:2018jnf}. It was found that the entropy of the massless scalar field in the interior of the black hole is proportional to the volume of the black hole and the variation rate of the interior entropy correlates with the variation rate of the Bekenstein-Hawking entropy by some simple relations. Specifically, for the Schwarzschild black hole, the ratio between the variation rate of the interior entropy and that of the Bekenstein-Hawking entropy in an infinitesimal Hawking radiation process is constant \cite{Zhang:2015gda,Zhang:2016sjy,Wang:2018dvo}.

Note that, previous studies neglected an important fact that thermodynamic quantities of black holes, such as entropy and temperature, will be corrected in the Hawking radiation due to quantum effect \cite{Zhang:2018nep}. One may wonder that whether the corrected Bekenstein-Hawking entropy and temperature of the black hole do change the result obtained before.  In this paper, we will introduce the corrected thermodynamics, the interior volume and interior entropy of the massless scalar field for the spherically symmetric black hole in Sec. \ref{rev}. We will generally calculate the ratio between the variation of the interior entropy and the variation of the Bekenstein-Hawking entropy for the spherically symmetric black hole in Sec. \ref{corr}. Lastly, remarks will be given in Sec. \ref{con}.

\section{Background setup}\label{rev}
\subsection{Corrected thermodynamics}
By investigating Hawking black body spectrum using tunneling mechanism, it was shown that the black hole possesses perfect black body spectrum as well as corrected temperature. At the first-order level in $\hbar$ (which has been set as 1), we have the corrected Hawking temperature for the black hole as \cite{Banerjee:2008ry,Banerjee:2009tz}
\begin{equation}\label{tempb}
T_{B}=T_{0}\left(1+\frac{\gamma}{2S_{0}}\right),
\end{equation}
where $T_{0}$ is the classical Hawking temperature defined through surface gravity $\kappa$ of the black hole as \cite{Hawking:1974sw,Bekenstein:1972tm}
\begin{equation}
T_{0}=\frac{\hbar\kappa}{2\pi ck_{B}},
\end{equation}
$S_{0}$ is the Bekenstein-Hawking entropy of the black hole which is found to be proportional to the event horizon area $A$ by \cite{Bekenstein:1973ur}
\begin{equation}
S_{0}=\frac{k_{B}c^{3}A}{4\hbar G}.
\end{equation}
Though lacking of an effective explanation for the Bekenstein-Hawking entropy, it has been shown that the renowned entropy-area law gets corrections from different perspectives, including the non-perturbative quantum gravity, the Cardy formula, the brick wall method, the tunneling method and many others \cite{Kaul:2000kf,Majumdar:2000pr,Cardy:1986ie,Kaul:1998xv,Carlip:2000nv,Govindarajan:2001ee,Mann:1997hm,Solodukhin:1997yy,Faizal:2014tea,Ali:2012mt,Sen:2012dw}. The first-order corrected entropy with logarithmic term can be expressed as
\begin{equation}
S_{B}=S_{0}-\frac{\gamma}{2}\ln S_{0}.
\end{equation}
It can also be obtained from (\ref{tempb}) by
\begin{equation}
S_{B}=\int\frac{dm}{T_{B}}.
\end{equation}
The first law of the black hole with corrected entropy and temperature can be written as \cite{Zhang:2018nep,Banerjee:2008ry}
\begin{equation}\label{firstlaw}
dm=T_{B}dS_{B}+\sum_{i}Y_{i}dX_{i},
\end{equation}
where thermodynamic quantities $X_{i}$ (such as the electric charge $q$) and $Y_{i}$ (such as the electric potential $\phi$) are individually the generalized displacements and generalized forces of the black hole.

\subsection{Interior volume and entropy of black holes}
In the flat spacetime described by the Minkowski line element
\begin{equation}
ds^{2}=-dt^{2}+dr^{2}+r^{2}d\theta^{2}+r^{2}\sin^{2}\theta^{2}d\phi^{2},
\end{equation}
the volume $V_{S}$ at time $t=t(r)$ possessed by the three-dimensional spacelike surface $\Sigma$ bounded by a two-dimensional surface $\mathcal{S}$ defined at $t=0, r=R$ can be written as
\begin{equation}
V_{\mathcal{S}}=\int_{0}^{R}4\pi r^{2}\sqrt{1-\left(\frac{dt(r)}{dr}\right)^{2}}dr,
\end{equation}
where we have set $t(R)=0$. We can see that the everyday volume $4\pi R^{3}/3$ can be obtained by two ways, {\it{i.e.,}} letting $t(r)=constant$ or finding the extreme value of $V_{S}$. These two ways are equivalent \cite{Christodoulou:2014yia}.

The four-dimensional spherically symmetric curved spacetime can be described by the line element
\begin{equation}
ds^{2}=-f(r)dt^{2}+g(r)dr^{2}+r^{2}d\theta^{2}+r^{2}\sin^{2}\theta^{2}d\phi^{2}.
\end{equation}
Using the Eddington-Finkelstein coordinate transformation
 \begin{equation}
 v=t+\int\sqrt{g(r)f^{-1}(r)}dr,
 \end{equation}
the spherically symmetric curved spacetime metric can be transformed to
\begin{equation}\label{ief}
\begin{aligned}
ds^{2}=&-f(r)dv^{2}+2\sqrt{f(r)g(r)}dvdr+r^{2}d\theta^{2}\\&+r^{2}\sin^{2}\theta^{2}d\phi^{2}.
\end{aligned}
\end{equation}
 In order to calculate the volume of a black hole bounded by a null event horizon, we can parameterize the coordinates $v, r$ in (\ref{ief}) using a dimensionless parameter $\lambda$. Then the line element (\ref{ief}) can be rewritten as
 \begin{equation}\label{iefx}
 \begin{aligned}
ds_{\Sigma}^{2}=&\left[2\sqrt{f(r)g(r)}\frac{dv(\lambda)}{d\lambda}\frac{dr(\lambda)}{d\lambda}-f(r)\left(\frac{dv(\lambda)}{d\lambda}\right)^{2}\right]d\lambda^{2}\\&+r^{2}d\Omega^{2},
\end{aligned}
\end{equation}
where $d\Omega^{2}$ is the line element of a two-dimensional sphere.

After that, the proper volume $V_{\Sigma}$ of the three-dimensional spherically symmetric surface bounded by the two-dimensional event horizon of the black hole can be got as
  \begin{equation}\label{func}
 \begin{aligned}
 V_{\Sigma}=&4\pi\int_{\lambda_i}^{\lambda_{f}}r^{2} d\lambda  \\&\times\sqrt{2\sqrt{f(r)g(r)}\frac{dv(\lambda)}{d\lambda}\frac{dr(\lambda)}{d\lambda}-f(r)\left[\frac{dv(\lambda)}{d\lambda}\right]^{2}},
 \end{aligned}
 \end{equation}
 where $r(\lambda_{i})$ and $r(\lambda_{f})$ respectively correspond to the outer event horizon and the inner horizon (for Schwarzschild black hole, the inner one degenerates to the central singularity point), $v(\lambda=\lambda_i )$ and $v(\lambda=\lambda_{f})$ respectively correspond to the beginning and ending of the collapse. The concept of simultaneity surface does not work in curved spacetime any more, so we are left with finding the extreme value of $V_{\Sigma}$ to obtain the interior volume of the black hole. Using Lagrangian method, it was found that the functional (\ref{func}) can be extremal when the coordinate $r$ of the two-dimensional surface $S$ satisfies \cite{Christodoulou:2014yia,Zhang:2019pzd}
 \begin{equation}
 \frac{dr(\lambda)}{d\lambda}=0.
 \end{equation}
 Then the interior volume of the spherically symmetric black hole in Einstein gravity is verified to be 
\begin{equation}\label{vsig}
\begin{aligned}
V_{\Sigma}&=\left[4\pi\int_{v(\lambda_i )}^{v(\lambda_{f})}r^{2}\sqrt{-f(r)}dv\right]_{max}\\&=4\pi v\left[r^{2}\sqrt{-f(r)}\right]_{max}\\&=H(m)v,
\end{aligned}
\end{equation}
where $[]_{max}$ means the extreme value of the expression, $m$ is energy of the collapsed object (mass of the black hole). We have set $v(\lambda_i )=0$ and $v(\lambda_f )=v$.

In the interior of the black hole, by taking a proper coordinate transformation \cite{Zhang:2015gda}
\begin{equation}
dv=-\frac{d\tau}{\sqrt{-f(r)}}+d\lambda,~dr=-\frac{d\tau}{\sqrt{-g(r)}}
\end{equation}
on the line element $ds_{\Sigma}^{2}$, where $\tau$ denotes the time in the interior of the black hole, we can consider statistical entropy of the massless scalar field  whose motion satisfies Klein-Gordon equation. Calculations give the statistical entropy of the massless scalar field in the interior of the four-dimensional spherically symmetric black hole expressed as \cite{Zhang:2015gda,Wang:2018dvo} 
\begin{equation}\label{ssig}
S_{\Sigma}=\frac{\pi^{2}}{45}T^{3}V_\Sigma.
\end{equation}

\section{The corrected thermodynamics and the interior entropy for black holes}\label{corr}
The volume (\ref{vsig}) of the black hole is an asymptotic expression at the late time $v\gg m$. Also, we have set the retarded time when the event horizon of the black hole forms as $v(\lambda_i )=0$. In the process of calculating the black hole volume (\ref{vsig}), we only considered the collapsed matter without Hawking radiation. Here, we aim to consider the effect of the Hawking radiation, with which the corresponding metric of the black hole can be described by the Vaidya geometry \cite{Vaidya1951}. The volume of an evaporating black hole formed by a massive shell at the retarded time $v(\lambda_i )=0$, which in its large mass limit can be reduced to (\ref{vsig}), was exemplified in \cite{Christodoulou:2016tuu} by using uncharged Vaidya geometry. In what follows, we will consider the evaporating black hole with large mass.  The black hole in the Hawking radiation can be viewed as a black body. The mass loss rate of the black hole in the Hawking radiation can be measured by the Stefan-Boltzmann law \cite{Ong:2015dja,Ong:2016xcq}
\begin{equation}\label{sbl}
\dot{m}=-\alpha a\sigma T^{4},
\end{equation}
where $\alpha$ is the greybody factor, $a$ is a positive radiation constant related to the Stefan-Boltzmann constant, $\sigma$ is the geometric optic cross section of the black hole (see \ref{ap}), $T$ is the temperature of the black body and also the Hawking temperature tested by a stationary observer at infinity, $\phi$ is the electric potential, the dot means the derivative with respect to the time $v$. 

In \cite{Zhang:2015gda} {\it{et al}}, the temperature and entropy of the black hole are viewed as uncorrected ones, {\it{i.e.}},
\begin{equation}
T=T_{0},~S=S_{0}.
\end{equation}
However, when the Hawking radiation is considered, the Hawking temperature and the Bekenstein-Hawking entropy of the black hole will get modified. Thus, the temperature and Bekenstein-Hawking entropy of the black hole should be
\begin{equation}
T=T_{B},~S=S_{B}.
\end{equation} 
Then we can know from (\ref{sbl})  that in an infinitesimal process, where the interior of the black hole can be viewed as being in a thermal equilibrium state with the event horizon \cite{Wang:2018dvo}, the differential relationship between the infinitesimal time $v$ and the infinitesimal mass loss of the black hole in the Hawking radiation should be
\begin{equation}\label{lossrate}
dv=-\frac{dm}{\alpha a\sigma T_{B}^{4}}.
\end{equation}
Here, we have replaced the uncorrected temperature in (\ref{sbl}) directly by the corrected temperature $T_B$. The reason we can do this replacement is that the energy flux leaking from the black hole is \cite{Harlow:2014yka}
\begin{equation}
\frac{dm}{dt}=\sum_{l^\prime ,m^\prime}\int_0^\infty \frac{d\omega}{2\pi}\frac{\omega P_{abs}(\omega,l^{\prime})}{e^{\beta\omega}-1},
\end{equation}
where $l^\prime, m^\prime, \omega$ are spherical harmonic index, azimuthal harmonic index, free fields in the spacetime, respectively. $P_{abs}(\omega,l^{\prime})$ is the absorption probability of the field, $\beta$ is the inverse of the black hole temperature. One can prove that the mass loss rate calculated by this way is consistent with the Stefan-Boltzmann law.

According to (\ref{vsig}) and (\ref{ssig}), the variation of the interior volume and interior entropy in an infinitesimal evaporating process for the black hole can be written as
\begin{equation}\label{hawone}
\begin{aligned}
dV_{\Sigma}&=Hdv+\frac{\partial H}{\partial m} dm\\&=-\frac{dm}{\alpha a\sigma T_{B}^{4}}\left(H-\alpha a \sigma T_{B}^{4}\frac{\partial H}{\partial m}\right)
\end{aligned}
\end{equation}
and
\begin{equation}\label{hawtwo}
dS_{\Sigma}=\frac{\pi^{2}}{45}T_{B}^{3}dV_{\Sigma}+\frac{\pi^{2}}{15}T_{B}^{2}V_{\Sigma}dT_{B}.
\end{equation}
Here in (\ref{hawone}), we have considered the retarded time $v$ and the black hole mass $m$ as variations. The retarded time $v$ is in fact $v(\lambda_{f})$ which denotes the end of the collapse, contrasting with $v(\lambda_{i})$, which corresponds to the beginning of the collapse \cite{Christodoulou:2014yia,Zhang:2019pzd}. $v(\lambda_{i})$ is the time that the horizon forms (which has been set to be zero in above). As the collapse continues, $v$ increases over time. $dv$ means that we choose two spheres $\mathcal{S}_{U1}$ and $\mathcal{S}_{U2}$ (defined by $\lambda$ and $\lambda +d\lambda$), which bound a surface $\Sigma_{U}$ with volume $dV_{\Sigma}$, so we have the first term in the right hand side (r.h.s) of (\ref{hawone}). When the Hawking radiation is included, the variation of the mass should be considered, so we have the second term in the r.h.s of (\ref{hawone}).
Note that here we also have replaced the uncorrected temperature $T$ in (\ref{ssig}) with $T_B$, as $S_\Sigma$ is Boltzmann statistical entropy and the black hole is supposed to be in a thermodynamic equilibrium  state with Hawking temperature.

Then we can have the infinitesimal variation of the interior entropy of the massless scalar field as
\begin{equation}\label{120918}
\begin{aligned}
dS_{\Sigma}&\sim \frac{\pi^{2}}{45}T_{B}^{3}dV_{\Sigma}\\&=-\frac{\pi^{2}}{45}\frac{dS_{B}}{\alpha a\sigma}\left(H-\alpha a \sigma T_{B}^{4}\frac{\partial H}{\partial m}\right),
\end{aligned}
\end{equation}
where we have used the assumption
\begin{equation}\label{condi}
\dot{T}_{B}\sim 0
\end{equation}
in the first line, which means that the radiation process is slow \cite{Wang:2019dpk,Wang:2018txl}, the first law of the black hole (\ref{firstlaw}) was considered in the second line. Note that the approximation here is only suitable for the black hole with large mass. When the Hawking radiation lasts for a long time (Page time), the mass of the black hole will reduce to be Planckian and these approximations may be invalid. Accordingly, we can know that the ratio $\rho_{c}$ between the infinitesimal variation of interior statistical entropy of the massless scalar field and the infinitesimal variation of the Bekenstein-Hawking entropy corresponding to the event horizon of the black hole in the Hawking radiation process can be expressed as
\begin{equation}\label{rho}
\begin{aligned}
\rho_{c}\equiv\frac{\dot{S}_{\Sigma}}{\dot{S}_{B}}&=-\frac{\pi^{2}}{45\alpha a\sigma}\left(H-\alpha a \sigma T_{B}^{4}\frac{\partial H}{\partial m}\right)\\&=-\frac{\pi^{2}H}{45\alpha a\sigma}+\frac{\pi^{2}T_{B}^{4}}{45}\frac{\partial H}{\partial m}.
\end{aligned}
\end{equation}
Obviously, one has
\begin{equation}
\rho_{c}>-\frac{\pi^{2}H}{45\alpha a\sigma}+\frac{\pi^{2}T_{0}^{4}}{45}\frac{\partial H}{\partial m}\equiv \rho_{0},
\end{equation}
where $\rho_{0}$ is the uncorrected ratio. Spherically symmetric black holes in Einstein gravity can be described by the metric in powers of $r^{-1}$ giving the leading terms as
\begin{equation}
f(r)\sim 1-\frac{2 m}{r}+\cdots,\,\, g(r)\sim 1+\frac{2 m}{r}+\cdots
\end{equation}
Then we can know
\begin{equation}\label{appro}
T_{B}\sim m^{-1}\sim T_{0},\,\,H\sim m^{2}.
\end{equation}
Moreover, according to (\ref{vsig}) and (\ref{si}), we can know that $H/\sigma$ in (\ref{rho}) is dimensionless. With these in mind, we have
\begin{equation}\label{res1}
0>\rho_{c}>\rho_{0}
\end{equation}
for large black hole mass $m$. Whereas for the small black hole mass $m$ ($m\sim 0$), we may have 
\begin{equation}\label{res2}
\rho_{c}>0>\rho_{0}\,\,  \text{or}\,\,  \rho_{c}>\rho_{0}>0.
\end{equation}
(\ref{res1}) means that the corrected ratio between the positive infinitesimal interior entropy variation of the massless scalar field and the negative Bekenstein-Hawking entropy variation in the Hawking radiation is greater than the uncorrected one in the regime of large black hole mass, as the first term in (\ref{rho}) plays the leading role. (\ref{res2}) means that the corrected ratio between the negative infinitesimal interior entropy variation of the massless scalar field and the negative Bekenstein-Hawking entropy variation in the Hawking radiation is also greater than the uncorrected one in the regime of small black hole mass, as the second term in (\ref{rho}) takes control.

\section{Remarks}\label{con}
According to (\ref{hawone}), (\ref{120918}) and (\ref{appro}), we have
\begin{equation}
\dot{V}_{\Sigma}>0, \dot{S}_{\Sigma}>0
\end{equation}
in the Hawking radiation process for black hole with large mass. In the evaporation, as Hawking radiation carries energy \cite{Taylor:1998dk,landsberg1989stefan}, we have
\begin{equation}
\dot{m}<0,
\end{equation}
then it is not difficult to obtain
\begin{equation}\label{reso}
\frac{\delta V_{\Sigma}}{\delta m}<0, \frac{\delta S_{\Sigma}}{\delta m}<0.
\end{equation}
We can know that during the Hawking radiation process, both the interior volume and interior Boltzmann entropy increase for black hole with large mass.

A black hole is usually viewed as a system containing $e^{S}$ independent states ($S$ is the Bekenstein-Hawking entropy), which, applying to the black hole with radiation, will naturally lead to a viewpoint that fewer and fewer states are available to be entangled with Hawking radiation when the black hole shrinks \cite{Page:1993wv}. According to (\ref{reso}), we can speculate that there are more and more volume and states (corresponding to the Boltzmann entropy) to correlate with Hawking radiation and  store up the information for the black holes. Thus, the information is not leaked out; conversely, it is hold by the increasing internal states \cite{Rovelli:2017mzl}.

Considering the corrected effect of Hawking radiation on the Hawking temperature and the Bekenstein-Hawking entropy of the black hole, we compared the Boltzmann statistical entropy variation inside the black hole with the Bekenstein-Hawking entropy variation in an infinitesimal evaporating process and found that
\begin{equation}
\rho_c >\rho_0.
\end{equation}
This result shows a simple algebraic relation between these two kinds of entropies, and corroborates the idea that information do have more internal space to be stored and information can be conserved even though we consider the quantum effects for the evaporating black holes with large mass. Thus, our result provides a new inspiration on the information conservation paradox during the black hole evaporation under the semi-classical quantum effects.

As pointed out in (\ref{res2}), the (un)corrected ratio we considered will not always be negative. This originates from (\ref{hawone}) which uncovers that small black hole mass will result in a positive volume variation. However, for these black hole with small mass ($m\sim 0$), it should be noticed that we can not use the corrected thermodynamics, as both the classical spacetime geometry and the equilibrium thermodynamics description for the Planckian scale black hole lose their effectiveness \cite{Sadeghi:2016dvc,Pourhassan:2017qxi}. Instead, non-equilibrium thermodynamics should be used to analyze the quantum black hole system \cite{Christodoulou:2016tuu,Astuti:2016dmk}. 

One should also note that our result is suitable for the Hawking radiation which takes charged particles away from charged black hole. To see this, one can use the generalized Stefan-Boltzmann law \cite{Wang:2018txl}
\begin{equation}
\dot{m}=-\sigma T^{4}A+\phi\dot{q},
\end{equation}
where $q$ is the electric charge and $\phi$ is the conjugated electric potential of the black hole. By taking the approximation $\dot{q}\sim 0$ in the evaporation, we can also get the ratio (\ref{rho}) between infinitesimal variation of the statistical entropy of the massless scalar field inside the black hole and infinitesimal variation of the Bekenstein-Hawking entropy.

\section*{Acknowledgements}
This work was supported by the Initial Research Foundation of Jiangxi Normal University.

\appendix
\section{Geometric optic cross section }\label{ap}

The radial effective potential for a null particle around a spherically symmetric black hole with $f(r)=1/g(r)$  is \cite{chandrasekhar1985mathematical}
\begin{equation}
V=\frac{1}{2}f(r)\frac{L^2}{r^2}.
\end{equation}
The radius $R$ of the unstable circular orbit for the null particle can be obtained from
\begin{equation}
Rf^{\prime}(R)=2f(R),
\end{equation}
which makes $V^\prime (R)=0$.
For the circular orbit of the null particle, we have
\begin{equation}
V=\frac{E^2}{2}.
\end{equation}
Then we can obtain the critical apparent impactor factor
\begin{equation}\label{si}
b_c^2 \equiv\frac{L^2}{E^2}=\frac{R^2}{f(R)}.
\end{equation}
Accordingly, the geometric optic capture cross section for the null particle is \cite{wald1984general}
\begin{equation}
\sigma= \pi b_c^2 =\frac{\pi R^2}{f(R)}.
\end{equation}
For the Schwarzschild black hole with
\begin{equation}
f(r)=g^{-1}(r)=1-\frac{2m}{r},
\end{equation}
we have
\begin{equation}
R=3m,~\sigma=27 \pi  m^2;
\end{equation}
For the charged Schwarzschild black hole with
\begin{equation}
f(r)=g^{-1}(r)=1-\frac{2m}{r}+\frac{q^2}{r^2},
\end{equation}
we have
\begin{equation}
R=\frac{1}{2} \left(\sqrt{9 m^2-8 q^2}+3 m\right)
\end{equation}
and
\begin{equation}
\sigma=\frac{\pi (s+3)^4 m^2}{8 \left(s-2 x^2+3\right)},
\end{equation}
where $x=q/m$ and $s=\sqrt{9-8 x^2}$.

\end{document}